\documentstyle[twocolumn,epsf,aps,prb]{revtex}
\begin{document}
\title{Phase Separation Models for Cuprate Stripe Arrays}

\author{R.S. Markiewicz and C. Kusko}
\address{Physics Department and Barnett Institute, 
Northeastern U.,
Boston MA 02115}
\maketitle
\begin{abstract}
An electronic phase separation model provides a natural explanation for a large 
variety of experimental results in the cuprates, including evidence for both 
stripes and larger domains, and a termination of the phase separation in the 
slightly overdoped regime, when the average hole density equals that on the 
charged stripes.  Several models are presented for charged stripes, 
showing how density waves, superconductivity, and strong correlations compete 
with quantum size effects (QSEs) in narrow stripes.  The energy bands associated
with the charged stripes develop in the middle of the Mott gap, and the 
splitting of these bands can be understood by considering the QSE on a single 
ladder.  
\end{abstract}

\narrowtext

\section{Introduction}

Possible electronic phase separation (EPS) in the cuprates has usually been 
found in terms of stripe phases.  Thus, neutron diffraction measurements find
evidence for fluctuating stripe order in La$_{2-x}$Sr$_x$CuO$_4$ (LSCO) 
associated with incommensurate inelastic neutron scattering\cite{Yam}, which can
be transformed into long range charged stripe order\cite{Tran} by co-doping 
with Nd or Eu.  Similar incommensurate peaks are found in other cuprates, and
long range charge ordered states are also found\cite{Mook} in strongly 
underdoped samples of YBa$_2$Cu$_3$O$_{7-\delta}$ (YBCO), with stripes parallel 
to the chains, while at higher doping short range stripe order is found at 
virtually all temperatures up to the pseudogap T$^*$.  However, EPS can also
manifest itself in the form of {\it larger domains}, particularly if the dopant
ions are somewhat mobile and can follow the hole motion.  These domains, long
known in La$_2$CuO$_{4+\delta}$ may recently have been found in scanning
tunneling microscope (STM)\cite{phassep} and microwave\cite{Oren} studies of 
Bi$_2$Sr$_2$CaCu$_2$O$_{8+\delta}$ (Bi2212).

Stripes are typically interpreted either in terms of charged domain walls in 
antiferromagnets or of restricted EPS.  While the models are virtually 
indistinguishible in the underdoped regime, they diverge as doping increases.  
A phase separation model implies the existence of a uniform end phase: at some 
doping antiferromagnetic (AFM) stripes must disappear, leaving a uniform state 
similar to that found on the charged stripes.  For slightly lower doping the 
AFM stripes form a minority -- a situation not possible in a domain wall 
picture.  (A note of caution: more recent domain wall models find a lower 
doping on the charged stripes, and as the doping decreases, the distinction 
between the two models gradually blurs.)
Evidence for the termination of the AFM 
stripes at a fixed doping\cite{Tal1} and recent observations of larger domain 
structures\cite{phassep} taken together provide extremely strong support for a 
phase separation scenario.

At this point it is essential to better characterize the nature of the stripes,
in particular the charged stripes, and to understand how their properties
affect physical phenomena, in particular photoemission 
spectra.\cite{SEK,Sei,OSP,ZEA,EOBB} 
While fluctuations\cite{SEK,Sei}  play an important role in the real cuprates, 
we have constructed ordered stripe arrays, for which detailed tight-binding 
calculations are possible\cite{OSP}.  Such calculations can aid in elucidating
the structure of the charged stripes, both for wide stripes -- what stabilizes 
the preferred hole density -- and for narrow stripes -- how do quantum size 
effects (QSE) modify properties of the stripes.  The present paper provides an 
extended analysis of these issues; some of these results were summarized
recently\cite{MK0}.

The paper is organized as follows.  Section II enumerates key issues which
must be addressed in any EPS model of stripes, including determination of
the hole density on charged stripes.  It is found that the model can 
simultaneously account for a large number of experimental results.  The charged 
stripes are probably stabilized by some competing order, either magnetic or 
paramagnetic (including charge density waves).  Section III shows that magnetic 
charged stripes can arise in a mean field Hubbard model, and can be either 
ferromagnetic or a linear antiferromagnetic (LAF) phase similar to the 
White-Scalapino stripes\cite{WhiSc}.  Calculations on
stripe arrays find that the charged stripes lead to midgap states near the Fermi
level; Section IV shows how these data can be interpreted in terms of quantum 
size effect (QSE) on single stripes.  Section V presents the results of single 
stripe calculations: competing charge density wave (CDW) and superconducting 
orders can exist on paramagnetic stripes, but they are strongly modified by the 
QSE. On LAF stripes, d-wave superconductivity and an unusual form of CDW are 
both found to persist down to the narrowest (2 cells wide) stripes.  Section VI 
includes results on stripe arrays: the LAF stripes produce
photoemission constant-energy maps in significantly better agreement with
experiment.  In the Discussion, Section VII, we summarize additional recent 
evidence which favors a phase separation model -- in particular, evidence that 
AFM stripes terminate slightly beyond optimal doping, where superconducting
properties remain strong, and that a second regime of phase separation exists
above optimal doping -- and we show that experiments are consistent with our
prediction that the superconducting gap grows in the underdoped regime.  
Conclusions are given in Section VIII.

\section{Key Issues for an EPS Model of Stripes}

A phase separation model of stripes is characterized by the two well-defined 
stable end phases between which phase separation takes place.  The insulating 
stripes are generally understood to be antiferromagnetic (AFM) -- essentially 
the same as the Mott insulator found in undoped cuprates.  The hole-doped 
stripes are assumed to have a finite doping, $x_0$.  This simple idea 
has three experimentally verifiable characteristic features: (i) {\it the stripe
phase must terminate} when $x=x_0$; (ii) there will be a {\it 
crossover} at a lower doping, $x_{cr}\sim x_0/2$ from magnetic-dominated 
($x<x_{cr}$) to charge-dominated ($x>x_{cr}$) stripe arrays; (iii) some {\it 
interaction} on the charged stripes stabilizes the end phase at $x_0$.  Here we
discuss current evidence for (a) the doping on the charged stripes, (b) evidence
for a crossover, and (c) the nature of the dominant interaction stabilizing the
charged stripes.  In addition, we ask what constraints the Yamada plot puts on
the model, and how a model based on EPS compares to a domain wall model.

\subsection{Hole Doping on Charged Stripes}

Recent evidence suggests that the stripes and pseudogap terminate at the same 
doping $x_0$ while superconductivity persists to higher doping\cite{Tal1}.  
However, the proper choice of $x_0$ requires some discussion.
The neutron diffraction measurements of Tranquada\cite{Tran} and 
Yamada\cite{Yam} have established that charged stripes in La$_{2-x}$Sr$_x$CuO$_
4$ (LSCO) have an invariant topology over the doping range $0.06\le x\le 0.125$,
acting as antiphase boundaries (APBs) for the AFM stripes and having a net 
doping of 0.5 holes per
stripe.  However, there are two models for how this charge is distributed: 
either in one row with average hole density 0.5 hole per copper site or in two 
rows with 0.25 hole per copper.  These two alternatives are often somewhat 
simplistically referred to as site order vs bond order.  The strongest evidence 
distinguishing between the alternatives comes from x-ray data on the charge 
order\cite{Zimm} at $x$=1/8: non-observation
of diffraction harmonics suggests a sinusoidal distribution of charge.  For a 
4-Cu repeat distance, two insulating and two charged rows would be exactly 
sinusoidal, whereas one charged and three insulating rows should have 
signifigant harmonic content.  A similar conclusion was reached by $\mu$SR
lineshape analysis\cite{Koj}.  However, the charge ordering peaks are weak, and 
it remains possible that fluctuations or disorder could wash out the harmonics. 

Direct evidence for the density on the charged stripes is found from 
low temperature NQR measurements\cite{TBG}, which find values $x_0\sim 0.18-0.19
$.  While this is close to the lower value, the small difference can also be
understood: this is a {\it local} measurement, and it is expected that some 
holes will be pushed off onto the magnetic stripes.
Indirect evidence favoring the lower hole density includes the fact that it is
easier to understand the properties of AFM stripes in terms of even-leg
ladders (e.g., the AFM stripes would be two coppers wide at 
$x$=1/8)\cite{Twer}, and that the stripe phase appears to terminate when the
average doping approaches $x$=0.25 (Section VII.A.1).  The lower doping is also 
more consistent with the tJ model simulations of White and 
Scalapino\cite{WhiSc}.

Tallon\cite{Tal0} finds an optimal doping at $x_{opt}=0.16$ for all cuprates, 
with respect to which the stripes terminate at a doping $x=0.19$.  However, it 
is hard to reconcile a common optimal doping with muon spin resonance 
data\cite{Uem0}, which find $T_c$ is optimized at very different values of $n_s
/m$ ($n_s$ is the pair density, which seems to scale with carrier 
concentration\cite{Talx1}, and $m$ effective mass) for 
LSCO and YBCO.  We assume instead that $x_{opt}$ 
scales with $n/m$, so if $x_{opt}=0.16$ for LSCO, it is 0.21 for YBCO, in good 
agreement with several estimates\cite{Talx}.  This also resolves a problem
with the thermopower.  While the thermopower appears to be universal for most
cuprates, and the best means of estimating the doping is from the room
temperature thermopower, LSCO is anomalous in that `overdoped' samples
still have high thermopower\cite{Talth}.  If the doping for YBCO is rescaled as
above, however, the thermopower data of LSCO fall on the universal curve.  
Hence, the anomaly for LSCO is not in the thermopower, but in a too low value of
$T_c$, which is accompanied by a too low value of $x_{opt}$.  It is likely that
these features are associated with a competing LTT phase, which is most
prominent in LSCO, and which also leads to the most nearly static stripe 
correlations.

Taking $x_{opt}$=0.21 for YBCO gives $x_0=19/16\times 0.21=0.25$, which we
believe holds for {\it all cuprates}, including LSCO, Section VII.A.1.  
This would lead to very wide charged stripes near optimal doping: the width of 
the charged stripes $N$ satisfies $N/(N+2)=16/19$, or $N=32/3\sim 10$ Cu wide.  
Hence, models of isolated quasi-one-dimensional charged stripes are likely to be
valid only in the far underdoped regime, while for the good superconductors a
better model would be a metal with intrinsic weak links\cite{ECJ1}.

\subsection{Crossover at 1/8 Doping}

For $x_0\sim 1/4$, the crossover $x_{cr}=x_0/2$ can be identified with the 1/8 
anomaly, where both charged and AFM stripes have their minimal width (2 Cu 
atoms).  There is considerable evidence that the doping 1/8, in addition to its 
special stability, acts as a crossover in the properties of the stripes.  Thus,
Uchida, et al.\cite{Uch}, studying the Hall coefficient $R_H$, find a 
crossover from one-dimensional behavior ($R_H\rightarrow 0$ as $T\rightarrow 0$)
for $x<x_{cr}=1/8$ to two dimensional behavior (coupled charged stripes) for
$x>x_{cr}$.  In YBCO, the spin gap grows slowly with doping for $x<x_{cr}$,
then more rapidly for $x>x_{cr}$; this behavior can be understood in terms of
coupled spin ladders, as the coupling changes with the width of the charged
stripes\cite{OSP}.  In Eu substituted LSCO\cite{Kat2}, the Meissner fraction
is negligibly small for $x<x_{cr}$, then grows roughly linearly with doping
until $x\simeq 0.18$, staying large up to at least $x=0.22$.  Finally, the 
two-magnon Raman peak in LSCO has a splitting at low temperatures which has been
associated with stripes\cite{SHay}, on analogy with similar observations in
La$_{2-x}$Sr$_x$NiO$_4$ (LSNO)\cite{LSNO}.  For $x<x_{cr}$ the ratio of the two 
peak frequencies is constant and consistent with a simple stripe model; for $x>x
_{cr}$ the lower frequency starts decreasing with doping.  Moreover, the higher 
frequency loses intensity with doping; near $x=0.26$, the {\it intensity} of one
mode approximately disappears, while the {\it frequency} of the other mode 
extrapolates to zero.

\subsection{Constraints from the Yamada Plot}

\begin{figure}
\leavevmode
   \epsfxsize=0.25\textwidth\epsfbox{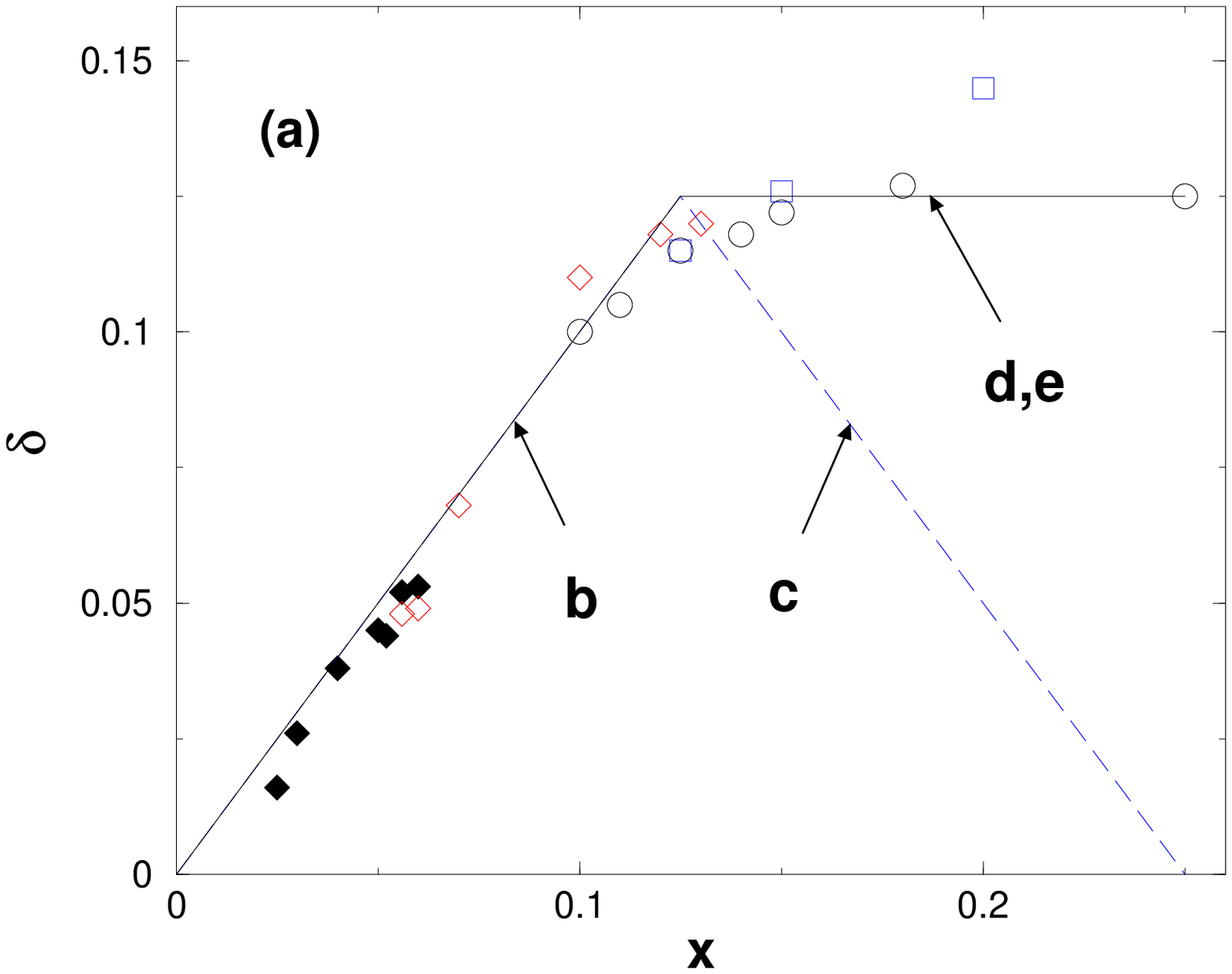}
\vskip0.5cm 
   \epsfxsize=0.48\textwidth\epsfbox{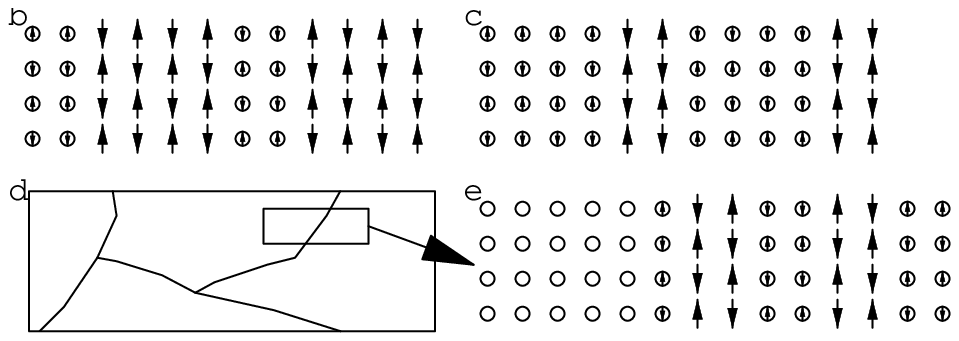}
\vskip0.5cm 
\caption{(a) Yamada plot of incommensurability $\delta$ vs doping $x$ for LSCO.
Open squares = elastic neutron scattering in Nd substituted 
samples\protect\cite{Tran}; others = inelastic neutron scattering for vertical
stripes (open circles \protect\cite{Yam} or diamonds \protect\cite{Waki}) or
diagonal stripes (filled  diamonds \protect\cite{Waki}). Dashed (solid) line =
prediction of EPS model without (with) commensurability effect at 1/8 doping.
(b,c) = Stripe phase model without commensurability effect, at $x$ = 1/12 (b)
and 1/6 (c).  For this figure, the charged stripes are assumed to have linear
antiferromagnetic order (Section III).  (d,e) = domains associated with 
commensurability pinning of 1/8 doped phase; (e) = blowup of (d).}
\label{fig:y1}
\end{figure}

The Yamada\cite{Yam,Tran,Waki} plot, Fig.~\ref{fig:y1}a, provides a severe 
constraint on any model of stripes: in LSCO the incommensurability $\delta$ is 
found to grow linearly with doping for $x<x_{cr}$ but to saturate for $x>x_{cr}
$.  Furthermore, the saturation value is just the incommensurability expected 
for $x=1/8$-doped stripes, $\delta_{sat}=x_{cr}$.  A similar saturation has been
reported in YBCO, but different groups find different values for $\delta_{sat}$:
$\sim 1/6$[\onlinecite{Ara1}] or $\sim 1/10$[\onlinecite{M4}].  If
one of these values proves correct, it would suggest some nonuniversality in 
$x_0$, perhaps associated with bilayer splitting.  In the domain wall model, 
the saturation is interpreted as evidence that the stripes stop 
changing in width, and additional holes leak into the antiferromagnetic 
background, washing out the stripes.  

For $x\le 1/8$, the EPS model agrees with the Yamada plot, Fig.~\ref{fig:y1}b:
increase in doping causes the AFM stripes to narrow, with no changes in the
charged stripes.  Note that, for concreteness we have assumed that the charged
stripes have LAF order (Section III); these stripes naturally act as antiphase 
boundaries (APBs), consistent with the neutron evidence.  However, the naive 
prediction of the EPS model is in disagreement with the Yamada plot for $x>1/8$,
Fig.~\ref{fig:y1}c: as the charge stripes get wider, the incommensurability 
should decrease, while the neutron peaks may broaden if the wider charge stripes
do not act as APBs. This behavior is not observed.  However, the model can be 
simply modified to explain the observed saturation, Fig.~\ref{fig:y1}d,e.  This 
would be a commensurability effect, with part of the sample pinned at 1/8 doping
while the rest forms a different phase (e.g., at 1/4 doping) where no stripes 
are present.  Such behavior is well known in nickelates, where coexistence of 
1/3 and 1/2 stripes is common.

Evidence for such commensurability effects can be found by comparing\cite{MK0}
the chemical potential $\mu$ in LSCO\cite{In1} and LSNO\cite{Sat}.  In both
materials, $\mu$ is approximately independent of doping between half filling 
($x=0$) and $x_{cr}$, with $x_{cr}=0.125$ in LSCO, 1/3 in LSNO, which has 
long-range charge order.  Remarkably, the simple prediction of a macroscopic 
phase separation model ($\mu$ constant throughout the phase separation regime)
is violated in LSNO: $\mu$ is {\it not} constant for $1/3<x<1/2$, although the 
stripe phase persists over the full doping range.  This is presumably a
commensurability effect, reflecting the coexistence of 1/3 and 1/2 stripe phases
for intermediate dopings.  The same doping dependence of $\mu$ is found in LSCO:
constant for $x<1/8$, variable for $x>1/8$.  Indeed, the LSNO data can be
scaled to that of the cuprates, giving a common doping dependence $\mu (x/x_c)$ 
in both compounds\cite{MK0}.  

\subsection{Domain Phase}

Recently local charged domains have been directly visualized by STM studies in 
Bi2212\cite{phassep} with gaps ranging from slightly overdoped to strongly 
underdoped.  Such behavior is extremely difficult to explain in the domain wall 
picture, but is easy to understand in terms of EPS, and indeed is consistent 
with the commensurability effects discussed above.  The gap distribution is
found to be broad rather than bimodal, but that is expected since EPS is very
sensitive to charged impurities, and the local gap will correlate with the local
density of dopant, presumably interstitial oxygen in Bi2212.  This sensitivity
to impurities leads to the question of which came first: is a preexisting EPS 
sensitive to local impurities, or does oxygen clustering\cite{Bal} provide the 
driving mechanism for domain formation?  Since the electronic inhomogeneity 
seems characteristic of most cuprates while there is considerable variety in the
doping counterions, the simpler interpretation would appear to be that the EPS 
is primary.  Thus, in La$_2$CuO$_{4+\delta}$ (LCO), the interstitial oxygens are
highly mobile, allowing the domains to grow to macroscopic size.  Similar 
clusters form in YBCO (here associated with chain oxygens), but can be 
suppressed in fully oxygenated samples by quenching\cite{Fis1}.  On the other
hand, well formed stripes appear when the doping counterions are least mobile, 
in LSCO.

The domains in Bi2212 would seem to be consistent, but a number of questions
remain about the role of annealing.  While these domains are regularly found in 
STM studies when the samples are cleaved at low-temperature, they can be 
annealed out at high temperatures\cite{Fis}.  This could be caused by a 
`melting' of the EPS, as is found in LCO near room temperature\cite{Ham}, and
can be tested by careful annealing studies.  Alternatively, it may be a question
of pinning the EPS.  Certainly, it is known that a weak domain disorder is 
necessary to pin vortices, to observe the vortex lattice via STM.\cite{Fis1}  
Also, the role of the interstitial oxygen in the superlattice modulations in 
Bi2212 needs to be better understood.  The strength of this modulation suggests
that it is associated with an ordering of the interstitial oxygen, but this 
should lead to a strong correlation between the domains and the superlattice, 
which seems not to be the case.  Further evidence for electronic inhomogeneity 
comes from microwave measurements: anomalies in Bi2212 have been interpreted in 
terms of similar electronic domains\cite{Oren}, suggesting that they are 
representative of the bulk, while measurements on other cuprates\cite{KusMa} 
find similar anomalous behavior which was interpreted in terms of pinned CDW's, 
possibly stripe related.  
A domain picture would also explain the persistence of nodal quasiparticles 
in the underdoped regime, at least down to 1/8 doping\cite{ZXdual}.  

\subsection{Comparison of Domain Wall Stripes and EPS Stripes}

Table I summarizes the discussion of the previous subsections, and compares the 
predictions of EPS models

\begin{tabular}{||c||c|c|c|c|c|c||}        
\multicolumn{7}{c}{{\bf Table I: Comparison of Stripe Models}} \\
            \hline\hline
Model & ${1\over 8}<x<{1\over 4}$&APB?&saturation &domains?&crossover&
termination \\
&&&at $x>{1\over 8}$?&&{\it at} $x={1\over 8}$?&at $x\sim{1\over 4}$?\\
    \hline\hline
Domain wall&fixed pattern&$\surd$&$\surd$&$\times$&$?\times$&$?\times$     \\
&holes spread out&&&&&        \\     \hline
EPS (simple)&charge stripes grow&$\sim$&$\times$&$\times$&$\surd$&$\surd$ \\    
&in width&&&&&          \\     \hline
EPS (commensurate &domain phase:&$\sim$&$\surd$&$\surd$&$\surd$&$\surd$  \\     
pinning)&$x={1\over 8}$: stripes&&&&&          \\    
&$x={1\over 4}$: uniform&&&&&          \\     \hline
\end{tabular}

\par\noindent
with domain wall models of stripes.  Domain wall 
models arose in early unrestricted Hartree-Fock (UHF) 
calculations\cite{HF,VLLGB,KMNF} for the doped tJ model, as heterogeneous ground
states in which the holes are confined to domain walls between AFM 
domains which act as APBs.  These domain walls are not driven by phase 
separation\cite{Cod}; the phase is realized as a long-period modulated AFM, with
holes doping the rows of spins where the moment changes sign\cite{IMac}.  When 
$t'=0$, there is one hole per row, a large hole doping which is inconsistent 
both with experiments on stripes in the cuprates, and with density matrix 
renormalization group (DMRG) calculations\cite{WhiSc}.  
More recent calculations reduce the doping, in ways that bring the model close
to the phase separation picture:  (1) assuming CDW order along the charged 
stripes\cite{CDW} lowers $x_0$ to 0.5, while (2) letting $t'<0$ results in 
$x_0\sim 0.2$.\cite{IMac}  Since the CDW order (case 1) lowers the free energy 
of the charged stripes, it should be possible to dope the system all the way to
a pure CDW at $x=0.5$.  In case 2, it is not explained why a low density 
appears; a good possibility is that this doping corresponds to the Van Hove
singularity (VHS), as in the phase separated stripes.  (This could be checked by
varying $t'$ and calculating $x_0(t')$.)  If such low-density domain walls do 
exist, the striped phase must somehow terminate when the average doping 
approached 0.2, but how this happens is not explained.  However, this 
calculation\cite{IMac} is not fully UHF, being restricted to periodic arrays, 
and UHF calculations\cite{MVz} find instead ferromagnetic charged stripes which 
can be understood within a phase separation model (see below).  

Since the stability of these $t'\ne 0$ domain wall stripes is unclear, in Table 
I a comparison is made with the earlier, higher-hole-density domain wall models.
For such high doping, stripe termination at $x=1/4$, or a major crossover near 
1/8, are both difficult to interpret.  Stripes as APBs are more natural in 
domain wall models, but we shall show that they can also arise in EPS models.
Note that the distinction between the models may be becoming moot, since in the
domain wall models there is now a search for a `hidden' order 
parameter\cite{CLaN}, and it is common for secondary order parameters to
be expressed predominantly on domain walls of the primary order\cite{KaVe}.

In conclusion, assumption of a charged stripe doping $x_0\simeq 0.25$ reconciles
the neutron diffraction data, evidence for a termination of the stripe phase
near $x_0$, and the 1/8 anomaly as a crossover effect near $x_0/2$, while
\vfill\eject
\par\noindent
commensurability effects can explain the saturation in the Yamada plot and the
STM observation of charge domains.  Further supportive evidence will be
presented in Section VII.A.  In the remainder of this paper, we will 
assume that $x_0=0.25$.  

\subsection{What Stabilizes Charged Stripes?}

In any phase separation model, a key issue is understanding the nature of
the charged stripes.  Indeed, since superconductivity seems to arise 
predominantly on these stripes, such understanding is likely to play an 
important role in elucidating the origin of the high superconducting transition 
temperatures.  For the stripe phase to exist, the doping $x_0$ must be 
particularly {\it stable}.  This can arise via an electronic {\it instability}, 
which opens up a gap over much of the Fermi surface, making the electronic 
phase nearly incompressible.  This `Stability from Instability' is a fairly 
general feature, underlying, e.g., Hume-Rothery alloys\cite{HumR}.  [This is a 
modification of an argument due to Anderson\cite{And}.]
Here, we explore a number of candidates for the predominant electronic
instability.  

In a related paper\cite{MarKII}, we will provide strong evidence that this 
`hidden order' is a form of CDW, which could include the flux phase.  However,
here we will explore a wider variety of possibilities.  One issue is that in the
low doping limit the charge stripes act as APBs for the AFM stripes.  Such an
effect arises naturally if the charged stripes have some residual magnetic
interaction, and we will explore this possibility.  However, in nickelates
charged stripes coupled to a CDW are found to act as APBs\cite{YYBiG}.  The 
large Hubbard on-site repulsion $U$ plays an important role.  Strong correlation
effects lead to two classes of charged stripe phases: 
either the constraint against double occupancy leads to magnetic order 
(magnetic charged stripes) or kinetic energy dominates, leading to a 
magnetically disordered phase (paramagnetic charged stripes), with correlations 
leading to reduced hopping, as in tJ\cite{Toh} and slave boson\cite{Pstr} 
models.  Section III will provide examples of both classes, denoted as Class B 
and Class A stripes, respectively.  Class A stripes could be simply correlated 
metals (as in tJ or slave boson calculations) or could have additional, e.g., 
CDW, order.  A crossover from magnetic to correlated paramagnetic groundstate 
arises as a function of doping in models of itinerant ferromagnets\cite{Faz}.  

For completeness, Class C stripes are defined as those arising not from phase 
separation but from long-range modulated AFM order (domain wall stripes).  Such 
stripes have been described in detail\cite{IMac}, and will not be further 
considered here.

In the next Section, we show that Class B (magnetic) charge stripes can arise
in a mean-field Hubbard model\cite{MarKI}, with the charged stripes displaying
either ferromagnetic (FM) or linear antiferromagnetic (LAF) (ordering vector 
$(\pi ,0)$) order.  The LAF stripes are very similar to White-Scalapino 
stripes\cite{WhiSc}.  The FM phase is stabilized by VHS nesting.  This FM phase 
may be present in ruthenates\cite{FMV}, but 
is unlikely to be relevant for the cuprates (for one thing, the FM stripes are 
likely to be diagonal, and do not form APBs\cite{MVz}, contrary 
to experiment).  We have suggested that other VHS-stabilized phases are more
likely\cite{Surv} (see also Ref.~\onlinecite{FriDi}), and here we explore the 
properties of a Class A CDW phase\cite{Pstr}. 
At a doping $x\sim 0.25$, the effects of correlations are relatively weak,
renormalizing the bandwidth by a factor of $\sim$2.  Thus for the present 
calculations on paramagnetic stripes, it will be assumed that renormalized 
parameters are used, and other effects of strong correlations will be neglected.

While LAF stripes are most stable when $t'=0$, we explore the possibility that 
they can be stabilized even when $t'\ne 0$ by on-stripe CDW or superconducting 
order.  Ordinarily, the CDW phase is believed to {\it compete} with strong 
correlation effects, but we find (Section V.B) that an unusual form of CDW phase
can {\it coexist} with LAF order: the charge density varies between zero and one
(not two) holes per atom.  

\subsection{Notation on Stripes}

In a stripe array, the alternating stripes are associated with the two stable
thermodynamic phases.  Here, we summarize the different ways these stripes are
denoted in this paper.  The stripe with lower hole doping is variously referred
to as `insulating' or `antiferromagnetic' (AFM).  (These are also known as 
magnetic stripes, since the charged stripes have considerably weaker magnetic 
order, but we will avoid that notation here.)
The stripe with higher hole doping is generically referred to as `charged' or
`hole-doped'.  At low temperatures, these stripes are also `superconducting', 
but at high temperatures, they are stabilized by some `hidden order', and one
purpose of this paper is to explore a number of possible orders.  The orders
fall into two groups: `magnetic charged' (Class B) stripes could have FM or LAF 
order (the latter are White-Scalapino-like stripes), while `paramagnetic' (Class
A) stripes could have CDW or flux-phase order.

\section{Phase Separation in a Mean-Field Hubbard Model}

Strong coupling models would seem to be natural for producing phase separated
or striped ground states.  Any magnetic ordering avoids double occupancy, while 
changing from one form of magnetic order to another, via, e.g., doping, requires
highly collective spin rotations, as competing orders are orthogonal.  While
superexchange leads to antiferromagnetic (AFM) insulators at half filling,
doping tends to favor textures with parallel spins (e.g., ferromagnetic (FM))
to maximize hole hopping.  Such ferron phases were introduced long before high
$T_c$\cite{Nag}, but it remains controversial whether such states are
ground states of the Hubbard model\cite{SuCCGB}.  While the tJ model does 
have phase separation for large $J/t$, it is unclear whether such phases extend 
to the values $J/t\sim 0.3$ expected for the cuprates\cite{WhiSc,PC,HelMan,pry}.

We have found phase separated solutions of the Hubbard model at mean-field
level\cite{MarKI}.  While these solutions are metastable in UHF calculations,
they closely resemble the WS stripes, and provide an interesting example of
phase-separation mediated stripe phases.  We find a well defined surface tension
for wide, isolated stripes, which decreases and changes sign as the stripes
become narrower.  When the surface tension becomes negative, the stripes no 
longer remain straight, and spontaneously meandering solutions are found.

These solutions are found by considering only low-order commensurate phases, 
with wave vector $q_x,q_y\sim$ 0 or $Q_i=\pi /a$ only.  The bare dispersion is
\begin{equation}
\epsilon_k=-2t(c_x+c_y)-4t'c_xc_y,
\label{eq:1}
\end{equation}
with $c_i=\cos{k_ia}$.  The Hubbard interaction $U\sum_i(n_{i\uparrow}-1/2)
(n_{i\downarrow}-1/2)$ leads to magnetic order with a mean-field magnetization 
$m_q$ at wave vector $\vec q$, and the quasiparticle dispersion becomes
\begin{equation}
E_{\pm}=\epsilon_+\pm E_0,
\label{eq:3x}
\end{equation}
where 
\begin{equation}
E_0=\sqrt{\epsilon_-^2+U^2m_q^2}
\label{eq:4x}
\end{equation}
and 
\begin{equation}
\epsilon_{\pm}={1\over 2}(\epsilon_k\pm \epsilon_{k+q})
\label{eq:2}
\end{equation}
For the cuprates, we expect\cite{OSP} $t\simeq 325meV$, $U/t\simeq 6$ and 
$t'/t\simeq -0.276$.  The magnetization is found self-consistently from
\begin{equation}
m_q=\sum_k (f(E_-)-f(E_+)){Um_q\over 2E_0},
\label{eq:6}
\end{equation}
with Fermi function $f(E)=1/(1+e^{(E-E_F)/k_BT})$.  The resulting free energy 
is
\begin{equation}
F=E_q-TS+U(m_q^2+{x^2\over 4}),
\label{eq:7}
\end{equation}
with
\begin{equation}
E_q=\sum_{k,i=\pm}E_if(E_i),
\label{eq:8}
\end{equation}
\begin{equation}
S=k_B\sum_{k,i=\pm}(f(E_i)ln(f(E_i))+(1-f(E_i))ln(1-
f(E_i))).
\label{eq:9}
\end{equation}
\begin{figure}
\leavevmode
   \epsfxsize=0.40\textwidth\epsfbox{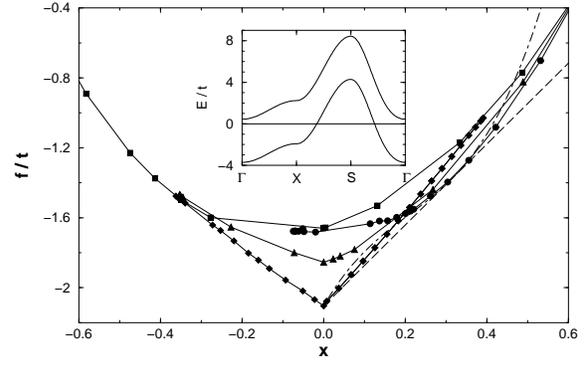}
\vskip0.5cm 
\caption{Free energy vs. doping for several magnetic phases of the Hubbard model
assuming $U=6.03t$, and $t'=-0.276t$.  Diamonds = AFM, 
triangles = LAF, circles = FM, and squares = PM phase.  Dashed lines = tangent 
construction; dot-dashed line = Eq. \protect\ref{eq:12}.  Inset: 
Dispersion of FM phase at $x=0.31$; Brillouin zone points 
$\Gamma =(0,0)$, $X=(\pi ,0)$, $S=(\pi ,\pi$).}
\label{fig:10}
\end{figure}

The competing phases include AFM for $\vec q=\vec Q\equiv (\pi ,\pi )$, 
FM with $\vec q=(0,0)$, and a linear antiferromagnet (LAF) with $\vec q=(\pi 
,0)$.  When the LAF stripes are two cells wide, this LAF phase closely
resembles the White-Scalapino stripes, Fig.~\ref{fig:y1}e.  The AFM state has 
lowest free energy at 
half filling, but (for $t'=0$) is unstable for finite hole doping.  For $t'=0$, 
there is phase separation between the AFM and LAF phases, while for finite $t'$ 
the phase separation is between AFM and FM phases, Fig.~\ref{fig:10}.  (When
electron-phonon coupling is included, it is found that the FM phase is
unstable with respect to a CDW phase\cite{MarKII}.)

\begin{figure}
\leavevmode
   \epsfxsize=0.40\textwidth\epsfbox{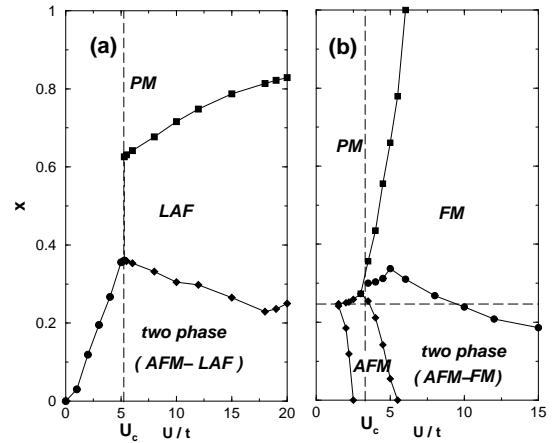}
\vskip0.5cm 
\caption{Phase diagrams, $x(U)$ for the Hubbard model, with $t'=0$ (a) or $t'=
-0.276t$ (b).  (Dashed line in (b) = doping of VHS.)}
\label{fig:6}
\end{figure}
The LAF stripes for $t'=0$ are discussed in Ref.~\onlinecite{MarKI}.  
For $t'\ne 0$ electron-hole symmetry is absent; for hole doping $x>0$, there is 
phase separation to a FM phase, consistent with recent simulations by 
Vozmediano, et al.\cite{MVz}, who find a uniform FM phase at $x=0.15$ for $U=8t
$, $t'=0.3t$.  However, {\it on the electron-doped side a uniform
AFM phase is stable over a large doping range}, suggestive of the asymmetry 
found in the cuprates.  The dot-dashed curve shows that in the phase separation 
regime, the low energy physics can be approximated by the form of free energy 
assumed in Ref.~\onlinecite{OSP}, 
\begin{equation}
f(x)=f_0x(1-{x\over x_0})^2,
\label{eq:12}
\end{equation}
(neglecting a term linear in $x$) with $x_0$ the hole doping of the uniform 
charged phase.
The FM phase is stabilized by VHS nesting, inset to Fig.~\ref{fig:10}, as found
previously\cite{FMV}.  The tangent construction tends to select the FM phase at 
dopings somewhat away from optimal nesting (Fermi energy above midgap).  Both 
regimes of phase separation seem to be driven by hole delocalization:
one-dimensional (along the LAF rows) when $t'=0$, two-dimensional for finite $t'
$.

The resulting phase diagrams $x$ vs $U$, Fig.~\ref{fig:6}, are strikingly 
different.  For $t'=0$, Fig.~\ref{fig:6}a, the phase separation is between the 
AFM and paramagnetic (LAF) phase for $U<U_c=5.3t$ ($U>U_c$), while 
for finite $t'$, Fig.~\ref{fig:6}b, there is generally a VHS-stabilized FM 
phase.  For small $U$ and $t'\ne 0$, there is a regime where simple spin-density
wave theory works and a uniform AFM phase is stable, but when $t'=0$ phase
separation persists for all finite $U$.  Note that $U_c$ marks a crossover 
between Class A (paramagnetic) and Class B (magnetic) charged stripes.  The 
value $U_c$ is close to the $U=6.03t$ expected in the cuprates, although for 
finite $t'$ $U_c$ decreases, $U_c\sim 3t$ for $t'=-0.276t$, and the range of $U$
for which paramagnetic stripes are stable becomes very small.
While these stripes are metastable in UHF calculations, we will show below
that the stability of phase separating stripes can be enhanced by 
{\it additional interactions} beyond the pure Hubbard model.

\section{Isolated Stripes vs. Arrays}

In ordered stripe arrays it is found\cite{OSP} that the AFM stripes have a
Mott-Hubbard gap, and the features near the Fermi level are associated with the 
charged stripes.  This charge stripe dispersion shows a series of 
quasi-one-dimensional features which qualitatively resemble the bands of an 
isolated stripe produced by QSE.  In this Section, we 
make a quantitative comparison with isolated stripes, and explore the mechanism 
of QSE-induced Van Hove splitting.  

\begin{figure}
\leavevmode
   \epsfxsize=0.40\textwidth\epsfbox{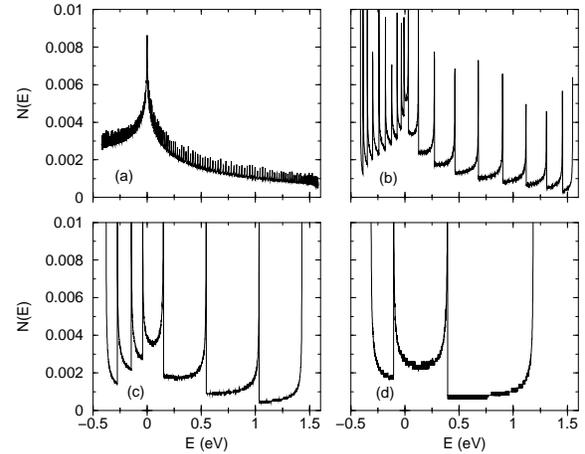}
\vskip0.5cm 
\caption{Density of states for a single stripe of width $N$ = 100 (a), 10 (b),
4 (c), or 2 (d) atoms. Based on Eq.~\protect\ref{eq:1} with t'/t=-0.276.}
\label{fig:1}
\end{figure}

For a single stripe $N$ Cu atoms wide, the dispersion is still given by 
Eq.~\ref{eq:1}, but the allowed $k_x$ values are quantized, with $k_x=k_m\equiv
m\pi /(N+1)$, $m=1,2,...,N$.  These are in fact the usual quantized Bloch bands,
but for large $N$ the quantization is not noticeable.  For small $N$, the 
dispersion appears as a series of $N$ overlapping one-dimensional (1d) 
dispersions.  Equation~\ref{eq:1} can be rewritten as $N$ 1d dispersions
\begin{equation}
\epsilon_{m,k_y}=-2tc_m-2(t+2t'c_m)c_y,
\label{eq:1e}
\end{equation}
These are the QSE, with corresponding density of states (dos) shown in 
Fig.~\ref{fig:1}.  Notice that for $N=100$, the VHS is readily apparent in the 
dispersion.  Even down to $N=2$, the VHS is clearly defined (albeit only within 
a finite interval) as the {\it locus of energies where all subbands overlap}.
In fact, {\it the QSE opens a gap at the VHS}, effectively lowering the kinetic 
energy of the electrons just like a conventional (e.g., CDW) gap.  The VHS
splitting can be found from Eq.~\ref{eq:1e}:
\begin{equation}
\Delta E_{VHS}=\epsilon_1^{max}-\epsilon_N^{min}=4t(1-c_1).
\label{eq:1f}
\end{equation}
This splitting has two
consequences: first, the VHS splitting enhances the stabilization of the
striped phase; but second, the QSE gap competes with other gaps, such as
CDW's and superconductivity.  However, while the QSE splits the VHS peak, 
substantial dos remains ungapped, so additional instabilities remain possible.
This competition will be discussed further in the next Section.

It should be noted that this is the first direct demonstration that the VHS can
be defined on a stripe only two atoms wide.  The definition is quite analogous
to the standard definition in two dimensions: the point at which the bands cross
over from electron-like to hole-like.

\begin{figure}
\leavevmode
   \epsfxsize=0.40\textwidth\epsfbox{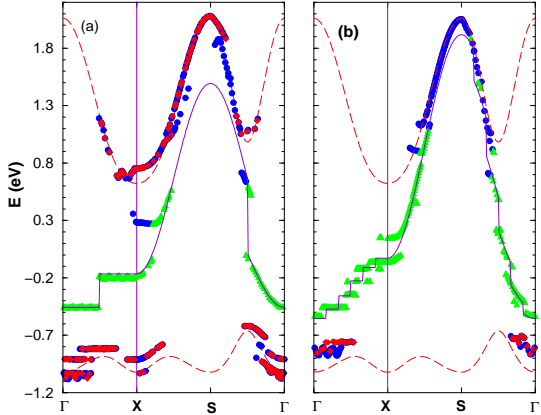}
\vskip0.5cm 
\caption{Dispersion of stripe arrays: (a) (6,2) array with charged stripes 2 Cu 
wide (AFM stripes 6 Cu wide); (b) (2,6) array with charged stripes 6 Cu 
wide (AFM stripes 2 Cu wide). Data from Figs. 7a and 7d respectively of 
Ref.\protect\onlinecite{OSP};
triangles (diamonds) = predominantly from charged (AFM) stripes, while
circles = mixed origin; dashed line = Mott bands of AFM stripes; solid
line = single (charged) stripe model, with $k_x$ approximated by nearest 
quantized value.}
\label{fig:1a}
\end{figure}

Figure~\ref{fig:1a} compares the gaps of a single stripe with those found in the
ordered stripe array\cite{OSP}; the array is labelled $(m,n)$ when the magnetic
stripes are $m$ coppers wide and the charged stripes are $n$ coppers wide.  In 
the array calculation, no competing order was introduced on the charged stripes, 
so the QSE provides the only gap.  Figure~\ref{fig:1a} shows that there is a 
very good match for both 2 Cu wide and 6 Cu wide stripes, although for the 2 Cu 
stripe, the VHS gap is somewhat larger for the single stripe than in the array. 
From Eq.~\ref{eq:1e}, the 1d bands have dos peaks at band bottom and top; the 
band bottom corresponds to $k_y=0$ -- i.e., the dispersion from $\Gamma
\rightarrow X$ is flat, at the energy corresponding to the lower dos peak.  
(The intensity along $\Gamma\rightarrow X$ is given by a structure factor, which
does not
directly come into a single stripe calculation.)  Along $X\rightarrow S$ one
should see the 1d dispersion extrapolating to the band top at $S=(\pi ,\pi )$.
Given the good agreement, it should be possible to analyze competing orders on 
a single stripe, for which the calculations are simpler (no need to 
self-consistently adjust doping on each row to account for charging effects).

\section{Ordering on Single Charged Stripes}

\subsection{CDW's and Superconductivity on Paramagnetic Stripes}

\subsubsection{Electron-Phonon Coupling}

In this section, we will develop two closely related models of the competition 
of CDW order and superconductivity {\it on a single charged stripe}.  
The first is a Class A paramagnetic stripe, stabilized by 
electron-phonon coupling\cite{MKK,BFal}, while the second has dominant 
electron-electron coupling, with (Class B) magnetic charged stripe order.

The earlier calculation\cite{MKK} used a van Hove stabilized CDW model
to describe the doping dependence of the pseudogap.  Here we reapply the model 
for a single stripe, by introducing the following modifications.  
(1) The calculation is carried out on a single, finite
width stripe.  (2) Pinning to the VHS arises naturally, since all stripes are at
the same doping.  (3) For closer approximation to experiment, d-wave 
superconductivity is assumed.  (4) Correlation effects due to Hubbard $U$ are 
neglected: previous slave boson calculations suggest that the main effect is a
bandwidth renormalization of a factor $\sim$2.\cite{Pstr} 

We briefly recall the energy dispersion and the gap equations of the 
model\cite{MKK}, generalized to d-wave.  In terms of a function
\begin{equation}
\Theta_{\vec k}=\cases{1,&if $|\epsilon_{\vec k}-\epsilon_F|<\hbar\omega_{ph}$;
                      \cr
                       0,&otherwise,\cr}
\label{eq:22}
\end{equation}
the gap functions are $\Delta_{\vec k}=\Delta_0\Theta_{\vec k}(c_x-c_y)/2$ for 
superconductivity, and $G_{\vec k}=G_0+G_1\Theta_{\vec k}\Theta_{\vec k+\vec Q}$
for the CDW.  The energy eigenvalues are $E_{\pm,k}$ and their negatives, with
\begin{equation}
E_{\pm,k}^2={1\over 2}(E_k^2+E_{k+Q}^2+2G_k^2\pm\hat E_k^2),
\label{eq:3}
\end{equation}
$E_k^2=\epsilon_k^2+\Delta_k^2$, $\hat E_k^4=(E_k^2-E_{k+Q}^2)^2+4G_k^2\tilde 
E_k^2$, $\tilde E_k^2=\epsilon_{k+}^2+(\Delta_k-\Delta_{k+Q})^2$,
$\epsilon_{k\pm}=\epsilon_k\pm\epsilon_
{k+Q}$, and the nesting vector $Q=(\pi ,\pi )$. 
The gap equations are
\begin{eqnarray}
\Delta=\lambda_{\Delta}\Delta\Sigma_{\vec k}{\Theta_{\vec k}(c_x-c_y)^2/4\over 
E_{+,k}^2-E_{-,k}^2}\times
\nonumber \\
\times\Bigl({E_{+,k}^2-\epsilon_
{\vec k+\vec Q}^2-\Theta_{\vec k+\vec Q}[\Delta_{\vec k}^2+G_{\vec k}^2]\over 2
E_{+,k}}\tanh{{E_{+,k}\over 2k_BT}}
\nonumber \\
-{E_{-,k}^2-\epsilon_
{\vec k+\vec Q}^2-\Theta_{\vec k+\vec Q}[\Delta_{\vec k}^2+G_{\vec k}^2]\over 2
E_{-,k}}\tanh{{E_{-,k}\over 2k_BT}}
\Bigr),
\label{eq:23}
\end{eqnarray}
\begin{eqnarray}
G_i=\lambda_G\Sigma_{\vec k}{\Theta_iG_{\vec k}\over E_{+,k}^2-E_{-,k}^2}\times
\nonumber \\
\times
\Bigl({E_{+,k}^2+\epsilon_{\vec k}\epsilon_{\vec k+\vec Q}-\Delta_{\vec k}^2-
G_{\vec k}^2\over 2E_{+,k}}\tanh{{E_{+,k}\over 2k_BT}}
\nonumber \\
-{E_{-,k}^2+\epsilon_{\vec k}\epsilon_{\vec k+\vec Q}-\Delta_{\vec k}^2-
G_{\vec k}^2\over 2E_{-,k}}\tanh{{E_{-,k}\over 2k_BT}}\Bigr),
\label{eq:24}
\end{eqnarray}
with interaction energies $\lambda_{\Delta}$ and $\lambda_G$, and 
$\Theta_0=\Theta_{\vec k}\Theta_{\vec k+\vec Q}$, $\Theta_1=1$.  

\begin{figure}
\leavevmode
   \epsfxsize=0.40\textwidth\epsfbox{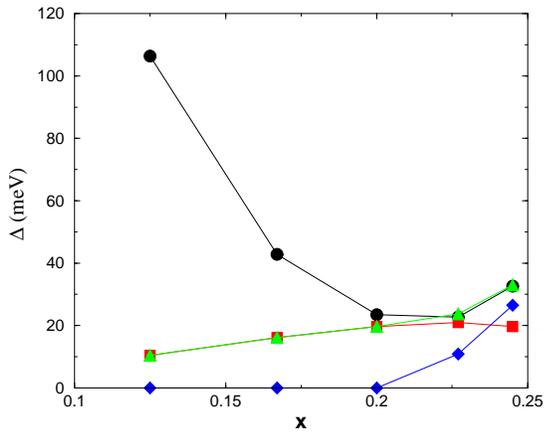}
\vskip0.5cm 
\caption{Gaps on a paramagnetic stripe, as a function of doping (equivalently: 
stripe width), assuming $x_0=0.25$, $\lambda_{CDW}=0.5eV$, $\lambda_{\Delta}=
0.25eV$.  Squares = s-wave superconducting gap; diamonds = CDW gap; triangles = 
combined gap; circles = total gap at ($\pi ,0$), including the quantum 
confinement gap.}
\label{fig:19}
\end{figure}

The CDW-superconducting competition was studied in bulk in 
Ref.~\onlinecite{MKK}.  The previous results are recovered for a sufficiently
wide stripe ($\sim$100 Cu wide).  For narrower stripes, 
it is found that the quantum confinement gap severely interferes with 
alternative gap formation.  Figure~\ref{fig:19} illustrates how the various 
gaps vary with stripe width, near $x_0=0.25$.  The data are plotted vs average 
doping, assuming a regular stripe array with 2-Cu wide AFM stripes (which
do not contribute to the gaps) and N-Cu wide charged stripes of doping $x_0$, 
giving an average hole doping $x=Nx_0/(N+2)$.  

The CDW gap is most sensitive to 
stripe width, but in the narrowest stripes superconductivity is also suppressed 
(the suppression is stronger for a d-wave gap).  
Strong instabilities are possible on the stripes, but they are shifted in 
doping away from $x_0=0.25$.  Thus, the
CDW instability requires both $E_{\vec k}$ and $E_{\vec K+\vec Q}$ to be near
the Fermi level; for a two-cell wide stripe, this is only possible near $x=0$.
On the other hand, superconductivity is possible anywhere, if the coupling is
strong enough.  For a two-cell wide stripe, the optimal superconductivity arises
when the Fermi level is at the {\it one-dimensional VHS} at the edge of one of
the stripe subbands.  This depends on $t'$, and for $t'=-0.276t$ falls at $x=
0.582$ (a larger gap is found on the electron-doped side, $x=-0.376$).
It should be noted that, even though the superconductivity is assumed to be
d-wave, in general a finite minimum gap is found on the stripe, even when the
CDW gap is zero.  This is because the vanishing d-wave gap can be sampled only
when the point $(\pi /2,\pi /2)$ is sufficiently close to the Fermi surface,
which in general requires $N$ to be odd (recall that the allowed values of $k_x$
are integer multiples of $\pi /(N+1)$) or very large.

\subsubsection{Electron-Electron Coupling}

In the above calculations, the $\lambda$'s arose from electron-phonon
coupling\cite{BFal}.  Similar contributions follow from electron-electron
coupling, in an extended Hubbard model.  For instance, the near-neighbor 
Coulomb repulsion has the following mean-field expansion:

\begin{eqnarray}
V\sum_{<i,j>,\sigma ,\sigma '}n_{i,\sigma}n_{j,\sigma '}=4V\sum_{\vec k,\sigma}
c^{\dagger}_{\vec k,\sigma}c_{\vec k,\sigma}
\nonumber  \\
-2V<O_n>\sum_{\vec k,\sigma}(c_x+c_y)c^{\dagger}_{\vec k,
\sigma}c_{\vec k,\sigma}
\nonumber \\
-8V<T_x>\sum_{\vec k,\sigma}c^{\dagger}_{\vec k+\vec Q,\sigma}c_{\vec k,\sigma}
\nonumber \\
-4V<T_y>\sum_{\vec k,\sigma}\tilde\gamma_kc^{\dagger}_{\vec k,\sigma}c_{\vec k,
\sigma}
\nonumber \\
+4V<T_z>i\sum_{\vec k,\sigma}\tilde\gamma_kc^{\dagger}_{\vec k+\vec Q,\sigma}
c_{\vec k,\sigma}
\nonumber \\
+4V\sum_{\vec k}\tilde\gamma_k(\Delta c^{\dagger}_{\vec k,\uparrow}c^{\dagger}
_{-\vec k,\downarrow}+\Delta^* c_{-\vec k,\downarrow}c_{\vec k,\uparrow})
\nonumber \\
+4NV(|\Delta |^2+<O_n>^2+<T_x>^2+<T>^2),
\label{eq:24b}
\end{eqnarray}
where $<T>^2=<T_x>^2+<T_y>^2+<T_z>^2$ and $\tilde\gamma_k=(c_x-c_y)/2$.
The first two terms in Eq. \ref{eq:24b} renormalize the chemical potential and 
the hopping $t$ respectively, and can be neglected.  The terms in $<T_i>$ 
comprise a pseudospin triplet of CDW-like distortions, with $T_x$ representing a
CDW similar to the one discussed above, $T_y$ being related to the 
low-temperature tetragonal distortion, and $T_z$ an orbital antiferromagnet,
closely related to the flux phase.  The remaining terms are a d-wave
superconductor.  The coefficients of the terms must be found self-consistently
by solving the gap equations:

\begin{equation}
\sum_{\sigma}<c^{\dagger}_{i,\sigma}c_{i,\sigma}>=1-x+2(-1)^{\vec r_i}<T_x>,
\label{eq:31}
\end{equation}
\begin{equation}
Im<c^{\dagger}_{i,\sigma}c_{i+\hat x,\sigma}>=
-Im<c^{\dagger}_{i,\sigma}c_{i+\hat y,\sigma}>=(-1)^{\vec r_i}<T_z>,
\label{eq:32}
\end{equation}
\begin{eqnarray}
Re<c^{\dagger}_{i,\sigma}c_{i+\hat x,\sigma}>=<O_n>+<T_y>
\nonumber \\
Re<c^{\dagger}_{i,\sigma}c_{i+\hat y,\sigma}>=<O_n>-<T_y>,
\label{eq:33}
\end{eqnarray}
\begin{equation}
<c^{\dagger}_{i\uparrow}c^{\dagger}_{i+\hat x,\downarrow}>=\Delta .
\label{eq:34}
\end{equation}
The terms $<O_n>$ and $<T_y>$ have recently been discussed by Valenzuela and 
Vozmediano\cite{VV}.  A detailed discussion of the competition between the
three CDW-like modes is given in Ref. \onlinecite{MarKII}.

\subsubsection{d-wave Superconductivity}

Retaining only the superconducting term in Eq.~\ref{eq:24b}, the interaction can
be derived from a quartic term
\begin{equation}
H'=\sum_{\vec k,\vec l}V_{\vec k,\vec l}c^{\dagger}_{\vec k,\uparrow}c^{\dagger}
_{-\vec k,\downarrow}c_{-\vec l,\downarrow}c_{\vec l,\uparrow},
\label{eq:35}
\end{equation}
with $V_{\vec k,\vec l}=2V(\cos{(k_x-l_x)a}+\cos{(k_y-l_y)a})$.  Assuming 
$\Delta_{\vec k}=\Delta_x\cos{k_xa}+\Delta_y\cos{k_ya}$, the gap equations can
be written in the form
\begin{equation}
\Delta_i=\sum_{j=x,y}A_{i,j}\Delta_j
\label{eq:36}
\end{equation}
($i=x,y$), with
\begin{equation}
A_{i,j}=-2V\sum_{\vec l}\cos{l_ia}\cos{l_ja}{\tanh{E_{\vec l}/2k_BT}\over 2E_{
\vec l}},
\label{eq:37}
\end{equation}
and $E_{\vec l}=\sqrt{(\epsilon_{\vec l}-e_F)^2+\Delta_{\vec l}^2}$.  

For the uniform charged state (infinitely wide stripe) $A_{x,x}=A_{y,y}$, and 
the gap symmetry can be simply analyzed.  The symmetry can be either d-wave
($\Delta_y=-\Delta_x$) or extended s-wave ($\Delta_y=+\Delta_x$), with the
choice $A_{x,y}\Delta_x\Delta_y>0$ giving the largest gap.  The $A_{x,x}$ term 
is always BCS-like, having the opposite sign from $V$, while the $A_{x,y}$ term
has the opposite sign from $A_{x,x}$, the integral being dominated by the
regions near the VHS's.  Hence, there are two possibilities: (i) attractive 
($V<0$) d-wave superconductivity or (ii) repulsive ($V>0$) extended s-wave.  
However, the latter would require $|A_{x,y}|>|A_{x,x}|$, which does not arise in
the present model, so only case (i) is possible.  These considerations readily
generalize to a finite-width stripe, for which $A_{x,x}\ne A_{y,y}$.

Agterberg, et al.\cite{ABG} recently introduced a model for `exotic' 
superconductivity in multiband superconductors.  If the Fermi surface consists 
of several inequivalent but degenerate pockets, the order parameter can consist
of symmetry allowed superpositions of the order parameters of the individual
pockets.  Equation~\ref{eq:37} can be thought of as a form of exotic 
superconductivity, with the degenerate VHS's playing the role of hole pockets.

\subsection{Modifications due to Magnetic Order}

\begin{figure}
\leavevmode
   \epsfxsize=0.40\textwidth\epsfbox{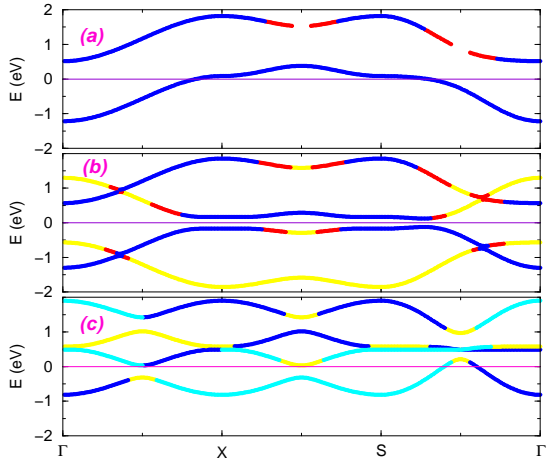}
\vskip0.5cm
\caption{Dispersion of linear antiferromagnetic (LAF) array along the
linear direction ($X$) (a), along with modifications due to d-wave
superconductivity (b) or CDW order (c).  $U/t$ = 6, $t'/t$ = 0, $V/t$ = 2 (b),  
0.1 (c).}
\label{fig:20a}
\end{figure}
In the above calculations, it was implicitly assumed that the doping is high
enough that the only role of the on-site repulsion $U$ is to renormalize the
band parameters.  However Baskaran\cite{Bas} recently estimated that near 
optimal doping correlation effects remain stronger than the kinetic energy 
associated with hopping.  Hence, it is important to look for stripe ground 
states which minimize this on-site repulsion (Class B stripes).  
The linear antiferromagnet (LAF) stripes discussed in Section III are a good
candidate for the cuprate charged stripes: they closely resemble the 
White-Scalapino stripes\cite{WhiSc}, have an appropriate doping, close to 
$x_0=0.25$, include strong correlations, and have the additional advantage that 
a two-cell wide LAF charge stripe acts as a natural APB for the 
AF stripes.  In this Section, we will explore these stripes, and show that they 
can be further stabilized by additional interactions.

\begin{figure}
\leavevmode
   \epsfxsize=0.40\textwidth\epsfbox{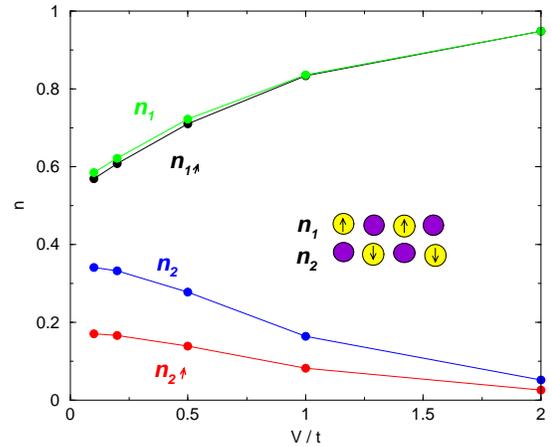}
\vskip0.5cm
\caption{Linear antiferromagnetic (LAF) array with CDW, showing spin and doping
distribution on different sites as a function of interaction strength $V$, with
$U/t$ = 6, $t'/t$ = -0.276.  Inset shows arrangement of atoms.}
\label{fig:20b}
\end{figure}

A special form of strongly correlated CDW is found to exist on a LAF.  The
charge and spin distribution is shown in the insert to Fig.~\ref{fig:20b}, with
the corresponding dispersion in Figure~\ref{fig:20a}c.  There is a strong
antiferromagnetic ordering on one sublattice, while most of the holes are 
confined on the other, nonmagnetic sublattice.  Whereas in a conventional CDW
the charge density is zero on one sublattice and two on the other, in this
strong coupling case the hole density varies from $~0$ to 1, and there is no
double occupancy, Fig.~\ref{fig:20b}.  Whereas the paramagnetic stripes 
were extremely sensitive to quantum confinement, these magnetic charged stripes 
are much less so:  this CDW is stable almost independently of the stripe 
width.  From Fig.~\ref{fig:20a}c, it can be seen that the gapped Fermi surface
still has hole pockets near $(\pi /2,\pi /2)$, which would lead to 
conducting stripes, consistent with optical properties\cite{RMopt}. 
However, it is only found near a hole doping $x=0.5$, and so does not
appear to be relevant for stripe physics in the cuprates.  

\begin{figure}
\leavevmode
   \epsfxsize=0.40\textwidth\epsfbox{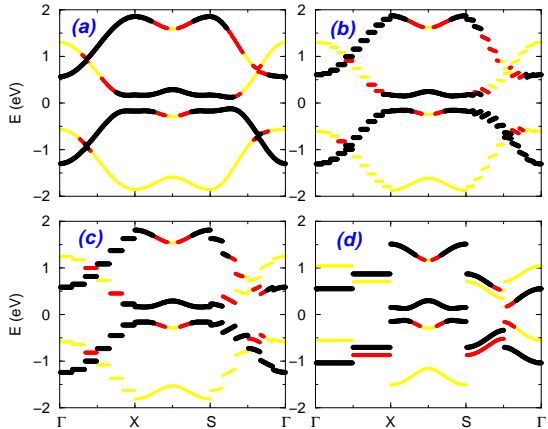}
\vskip0.5cm 
\caption{Dispersion of a LAF with d-wave superconducting order for a uniform
system (a) or a single stripe of width $N$ = 10 (b), 6 (c), or 2 (d) atoms.
Darkness of line reflects relative intensity of dispersion feature.}
\label{fig:20}
\end{figure}

\begin{figure}
\leavevmode
   \epsfxsize=0.40\textwidth\epsfbox{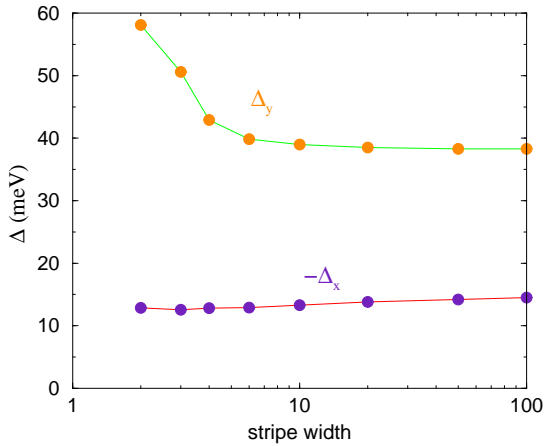}
\vskip0.5cm    
\caption{Linear antiferromagnetic (LAF) array with `d-wave' superconductivity, 
showing magnitude of gap along ($y$) or across ($x$) the stripes, as a function
of stripe width.}
\label{fig:20c}
\end{figure}

Away from this doping, CDW instabilities are relatively weak and it is possible 
to stabilize d-wave superconductivity, Figs.~\ref{fig:20a}b,\ref{fig:20}.  While
the overall dispersion varies with stripe width, the superconducting gap is also
relatively insensitive to the width, and actually increases for the narrowest
stripes, Fig.~\ref{fig:20c}.  Note that the order parameter is not a pure 
d-wave, the gap along the stripe being larger.  Such a large anisotropy is not
consistent with tunneling measurements of the gap; it is possible that the
anisotropy is reduced by strong interstripe coupling.  On the other hand, a 
large gap anisotropy has been found in YBCO\cite{YBFS}, where the stripes are
aligned along the chain direction\cite{Mook}.


Note in Figs.~\ref{fig:20a}b,~\ref{fig:20} that the combination of LAF and 
d-wave order leads to a finite minimum gap over the full Fermi surface.  While
the pure LAF phase is not stabilized by the VHS, the d-wave superconductivity
is optimized when the Fermi level is at the $(\pi ,0)$ VHS -- at essentially the
{\it same doping}, $x_0=0.245$, as the VHS on a paramagnetic stripe!

\begin{figure}
\leavevmode
   \epsfxsize=0.40\textwidth\epsfbox{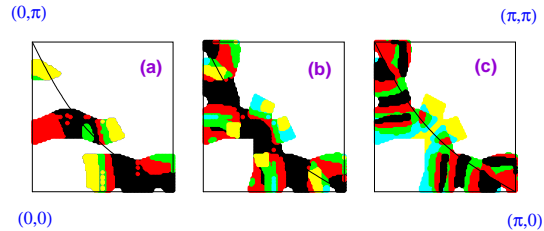}
\vskip0.5cm 
\caption{Constant energy cuts of photoemission dispersion for a (2,2) stripe
array, within 200 meV of the Fermi level.  Lines = Fermi surface of bulk (or 
very wide) charged stripes. Relative intensity increases with darker shading. 
(a) Representative of single domain sample; (b) for multidomain sample 
(symmetrized about the zone diagonal; (c) 
with diagonal-suppressing matrix element, $M=|c_x-c_y|$.}
\label{fig:21}
\end{figure}

\section{Extension to Arrays}

\begin{figure}
\leavevmode
   \epsfxsize=0.40\textwidth\epsfbox{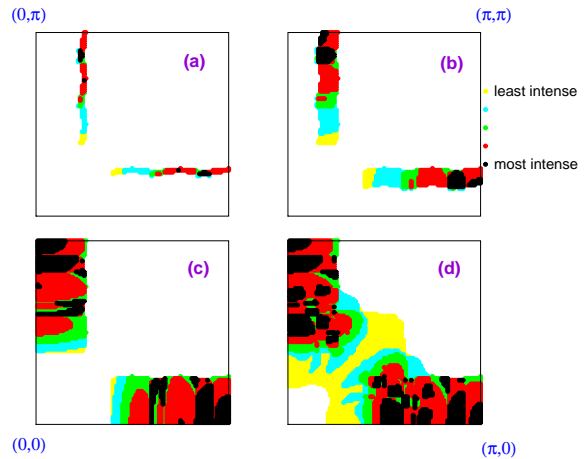}
\vskip0.5cm 
\caption{Constant energy cuts of photoemission dispersion for a (2,2) stripe
array, with LAF charged stripes, within (a) 30 (b) 100, (c) 200, or (d) 500 meV 
of the Fermi level.}
\label{fig:22}
\end{figure}

\begin{figure}
\leavevmode
   \epsfxsize=0.40\textwidth\epsfbox{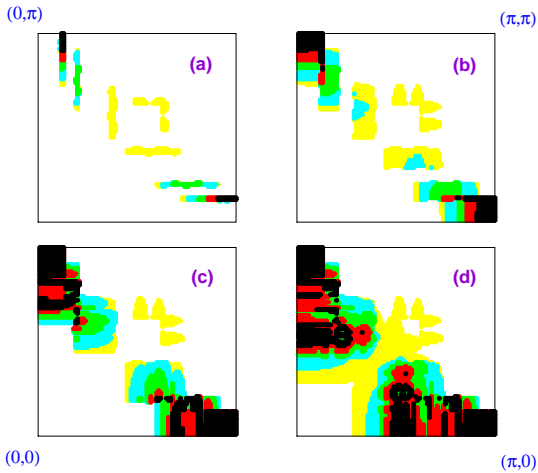}
\vskip0.5cm 
\caption{Constant energy cuts of photoemission dispersion for a (2,6) stripe
array, with LAF charged stripes, within (a) 30 (b) 100, (c) 200, or (d) 500 meV 
of the Fermi level.}
\label{fig:23}
\end{figure}

In Ref.~\onlinecite{OSP}, the Fermi surface was calculated for a series of 
ordered stripe arrays.  These results can now be compared to experimental
photoemission data\cite{ZZX}.  For this purpose we replot the data as integrated
spectral weight over a finite energy cut within energy $\Delta E$ of the Fermi
surface.  Figure~\ref{fig:21}a shows a cut with $\Delta E$ = 200meV, for the 
model of a 1/8 doped stripe array (i.e., $x=0.125$)\cite{OSP}.  The pattern is 
readily understood: the stripe superlattice leads to a number of 
quasi-one-dimensional bands; however, due to structure factor effects they have 
significant intensity only near the original Fermi surface, solid line in Fig. 
\ref{fig:21}a.  For comparison with experiment, the calculated spectral weight 
is symmetrized $(0,\pi )\leftrightarrow (\pi ,0)$ in Fig.~\ref{fig:21}b to 
represent a sample with regions of stripes running along both $X$ and $Y$ 
directions.  Finally, an empirical matrix element is included, 
Fig.~\ref{fig:21}c, which extinguishes spectral weight along the zone diagonal, 
$(0,0)\rightarrow (\pi ,\pi )$, similar to the matrix element assumed in 
analyzing Bi2212, Refs.~\onlinecite{ZZX,ZXX}.  The resulting Fermi surface
maps for several values of $\Delta E$ are illustrated in Ref. \onlinecite{MK0}
for paramagnetic charged stripes, and in Figs. \ref{fig:22} ($x=1/8$) and
\ref{fig:23} ($x=0.1875$) for LAF stripes.

For both models, the stripe band nearest $(\pi ,0)$ is in good agreement with 
experiment: there is little dispersion perpendicular to the stripe, while the 
intensity falls off toward $(0,0)$ due to the structure factor effect.  
In general, the LAF stripes are in better agreement with
experiment, since the additional subbands predicted for paramagnetic stripes 
(moving from $(\pi ,0)$ toward $(0,\pi )$) are not seen in the experiment. 
While the matrix element improves the agreement, theory suggests that this
effect is present only for certain photon polarizations\cite{LiBan}.  
One disagreement with experiment for both models is that for shallow energy 
cuts (30, 100 meV) the experiment still finds a smeared dispersion rather than 
a sharp Fermi surface.  This is presumably an effect of stripe fluctuations.  

It should be noted that all the spectral weight in 
Figs.~\ref{fig:21}-\ref{fig:23} is associated with the charged stripes; the 
lower Hubbard band of the AFM stripes lies below 0.5eV in LSCO.  It is
somewhat surprising that the spectral weight nearest the Fermi level is near
$(\pi ,0)$, since this is where the pseudogap arises.  Nevertheless, our
calculation reproduces both the (quantum confinement) pseudogap, 
Fig.~\ref{fig:1a}, and the spectral weight distribution.

\section{Discussion}

\subsection{Additional Evidence for Phase Separation}

\subsubsection{Termination of Stripes}

\begin{figure}
\leavevmode
   \epsfxsize=0.33\textwidth\epsfbox{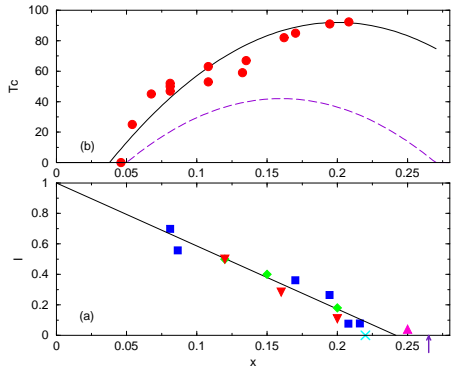}
\vskip0.5cm 
\caption{(a) Magnetic inelastic scattering intensity vs $x$ in 
YBCO (squares\protect\cite{Bourges}) and LSCO (diamonds\protect\cite{Tran} and
triangle\protect\cite{Yam}).  Inverted triangles = (scaled) effective exchange
constant in RE substituted LSCO, estimated from slowing of spin 
fluctuations\protect\cite{Hunt2}.
(b) Corresponding $T_c(x)$: solid line = YBCO (circles\protect\cite{Bourges}); 
dashed line = LSCO.}
\label{fig:3}
\end{figure}

The two most common interpretations for the pseudogap are in terms of either
precursor pairing or a competing order parameter, which may be 
magnetic\cite{KaSch,Zhang5} or charge-density wave (CDW)\cite{Pstr,RAK}.  We 
will argue in the next subsection that the latter possibility is more likely. 
Such competing instabilities arise naturally in a phase separation model (e.g.,
Refs. \onlinecite{CDC,OSP}), and we have long argued that the pseudogap is a
manifestation of local phase separation\cite{RM5}.  It should be noted that a
pseudogap arises in the nickelates\cite{Kats} in conjunction with stripe
fluctuations, and turns into a true gap at the charge ordering temperature.
Therefore, the fact that the pseudogap closes in the overdoped regime in the
cuprates strongly suggests that the stripe phase terminates at the 
same doping.  Direct evidence  for this has recently 
appeared\cite{OSP,Tal1,Bou,Mook}.  Figure \ref{fig:3}a compares the intensity 
$I$ of the inelastic neutron peaks near $(\pi ,\pi )$ as a function of doping 
in YBCO\cite{Bourges} and LSCO\cite{Tran,Yam}.  In a stripe picture, $I$ should 
be a measure of the fraction of
material in AFM stripes.  Remarkably, the intensity extrapolates to zero at
nearly the same doping in both materials, even though $T_c(x)$ peaks at 
substantially different dopings.  For both materials, this is the doping at 
which the pseudogap closes.  While in YBCO, these peaks have been interpreted as
a Fermi surface nesting effect\cite{nest1,nest2}, this cannot explain 
the charge order fluctuations, and in LSCO the Tranquada data\cite{Tran} is 
associated with {\it elastic} magnetic peaks.  Furthermore, a study of the 
neutron diffraction pair distribution function\cite{Boz} for LSCO finds 
evidence for charge fluctuations, presumably associated with stripes.  The 
excess fluctuations are maximal near $x=0.15$, and terminate near $x=0.25$.  
Strong reductions of thermal conductivity associated with stripe scattering 
also terminate at a comparable doping\cite{Bab}, while a recent optical 
study\cite{TiP} finds evidence for a quantum critical point at a similar doping,
$x\sim 0.22$.  Moreover, in Bi2212, Tokunaga, et al.\cite{TIYM} have introduced
a new crossover temperature $T_{mK}$ based on Cu NMR, below which AFM 
correlations develop; they find $T_{mK}\rightarrow 0$ near $x=0.26$.

A recent NQR study of the slowing of spin fluctuations in RE
substituted LSCO\cite{Hunt2} finds that the effective spin stiffness $\rho_s
^{eff}$ (or equivalently the effective exchange constant) scales to zero at a 
comparable doping; the inverted triangles in Fig.~\ref{fig:3} show $2\pi\rho_s
^{eff}/460K$.  As might have been anticipated from Sections II,IV, the doping
dependence of $\rho_s^{eff}$ changes radically below $x=0.12$.  Note that while
the integrated neutron intensity scales approximately with the area fraction of
charged stripes, $2\pi\rho_s^{eff}$ scales to $\sim$460K as $x\rightarrow 0$.
This is only 1/4 of the actual spin stiffness, $2\pi\rho_s=1.13J=1730K$ in the
undoped AFM.  The change by nearly a factor of four is suggestive of a 
dimensional reduction (lower coordination), but for an isolated straight spin
ladder, a factor of two might have been expected.

It is important to note the 
proximity of this termination of the phase separation regime to the VHS: the 
arrow in Fig.~\ref{fig:3}a shows the doping at which the pseudogap in the heat 
capacity\cite{Lor} closes, leaving an approximately logarithmic peak\cite{Surv},
while the $\times$ indicates the point at which photoemission finds the VHS
crossing the Fermi level\cite{ZXTex}.  Termination of the stripe phase close to 
a VHS is an important prediction of our EPS model of stripes.

Figure~\ref{fig:3} has a bearing on the discussion given earlier of optimal 
doping in YBCO, or equivalently of the doping $x$ at which the stripe phase
terminates.  Setting optimal doping at $x=0.16$ for all cuprates puts the
termination at doping $x_0=0.19$.  However, this value is clearly too low for 
LSCO: when the stripes are gone, magnetic correlations should be weak, but there
is evidence for long range magnetic order at $x=0.2$\cite{Tran} and $0.21$, the
latter coupled with a suppression of $T_c$\cite{Koi}.  On the other hand, both
facts are compatible with $x_0\sim 0.25$, Fig.~\ref{fig:3}.

The termination of the pseudogap has also been interpreted in terms of a 
quantum critical point (QCP)\cite{Varm,Sei}.  At first glance, it looks 
straightforward to distinguish a model of EPS, which is by nature first order, 
from a QCP, which is typically second order.
However, the phases are actually quite similar.  Due to strong coulomb effects, 
the EPS is restricted to nanoscopic scale, so that, e.g., the superconducting 
critical temperature $T_c$ changes smoothly with doping.  On the other hand, 
near a QCP with quenched disorder, long-time correlations can lead to effective 
spatial inhomogeneity\cite{BelKirv}.

\subsubsection{Doping Dependence of Charged Stripes and Superconductivity}

We suggested earlier\cite{OSP} that the peak and hump features seen in 
photoemission from Bi2212 were associated with the charged and the AFM 
stripes respectively.  As such, the intensity of the peak should have the
doping dependence predicted for charged stripes, with the intensity increasing 
from zero at half filling, approximately linearly with doping $x$.  This has 
now been verified experimentally\cite{ZXTex,DEW}.  Moreover, the maximum 
intensity of the spectral weight occurs at the same doping\cite{ZXTex} $x_0$ 
discussed above, where the stripe phase terminates.  Remarkably, the peak 
spectral weight closely tracks $T_c$, suggesting that {\it the superconducting 
pairs live on the charged stripes}, as predicted by several 
models\cite{Surv,MaB}.  Consistent with this, a number of measures of the
strength of superconductivity (condensation energy, critical current) are 
optimized at this same point\cite{ZXTex,Tal1,Tal2} where the charge stripe 
intensity is maximum and AFM stripes vanish.  The fact that $T_c$ itself is 
actually optimized at a slightly lower doping may be a hint that stripes can
enhance the superconducting gap, as found above, Fig.~\ref{fig:20c}.

The sharp falloff of spectral weight at higher doping may be an
indication for a {\it second regime of phase separation} in overdoped samples,
where the new phase is not intrinsically superconducting.  
Some additional evidence for such a phase separation is discussed below.

\subsubsection{Second Regime of Phase Separation}

It is important to note that as soon as the pseudogap closes, experiments
find a peak in the density of states consistent with that expected for a VHS.
This is discussed in Ref. \onlinecite{Surv} on p. 1203 (Fig. 21); more recent
evidence is in Ref. \onlinecite{TalN}.  For even higher dopings there are hints
of a new regime of phase separation.  This is an improtant prediction of the
model: if stripes are stabilized by lowering the electronic energy at a VHS,
then there could be a {\it second range of phase separation} in the overdoped
regime, with the non-superconducting phase totally unrelated to the AFM 
state at half filling.

Earlier experimental evidence for this second regime of phase separation in the
cuprates has been presented in References \onlinecite{Surv} (Section 11.6) and
\onlinecite{BGood} (see also Wen, et al\cite{Wen}).  Recently, Loram\cite{Lor} 
has presented a detailed 
analysis of the heat capacity of La$_{2-x}$Sr$_x$CuO$_4$ over an extended 
doping range.  In the range $0.26<x<0.3$, the temperature dependence of the
Somerfeld constant $\gamma$ is consistent with that expected for a uniform
phase close to a VHS (Ref. \onlinecite{Surv}, Fig. 21).  For lower doping, 
$\gamma$ falls off too rapidly to be associated with rigid-band filling away 
from a VHS, but can be easily explained in terms of the opening of a pseudogap.
For higher hole doping, $\gamma$ again falls off\cite{Lor2} too rapidly to be 
due to rigid band filling, although without an obvious pseudogap forming.  The 
near symmetry of $\gamma$ between the overdoped and underdoped sides is 
suggestive of two regimes of phase separation.  
Further possible evidence for this second regime comes from tunneling
studies\cite{Deut} which find a subdominant order parameter in optimally and
over-doped YBCO (i.e., in the same doping range where a split superconducting
transition is found\cite{Kald}).  This could arise from an 
Allender-Bray-Bardeen mechanism\cite{ABB}, with the d-wave superconducting
stripes inducing s-wave (or $d_{xy}$) superconductivity on adjacent 
normal-Fermi-metal stripes.  Also, microwave measurements\cite{Oren} find a
crossover near optimal doping from superconductor-insulator to
superconductor-metal domains.

Uemura\cite{Uem} has also recently proposed phase separation in the overdoped
cuprates.  In extending the Uemura plot to the overdoped regime, a `boomerang'
effect is found: both $T_c$ and $n_s/m$ decrease in parallel\cite{Uem2}.  Here $
n_s$ is the superfluid density and $m$ is the effective mass, and their ratio is
extracted from $\mu$SR measurements of the penetration depth.  Uemura\cite{Uem}
suggests that this proportionality between $T_c$ and $n_s/m$ can most easily be
accounted for in terms of (nanoscale) phase separation between the 
superconductor and a non-superconducting phase.  Thus for both overdoped and
underdoped cuprates\cite{MG}, the scaling between $T_c$ and $n_s$ can be 
understood in terms of phase separation.  This has important implications
for superconductivity: {\it superconductivity arises only at a unique 
composition near optimal doping.}  The present results suggest that even at this
doping, the superconductivity must compete with a second order parameter. 

\subsection{Superconducting Fluctuations}

\begin{figure}
\leavevmode
   \epsfxsize=0.33\textwidth\epsfbox{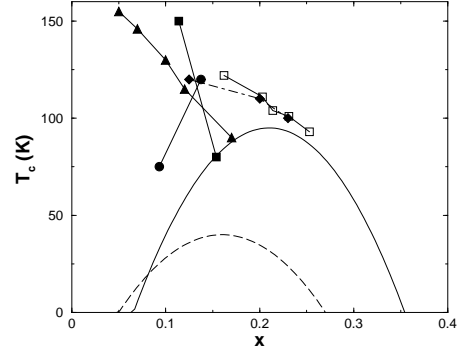}
\vskip0.5cm 
\caption{Doping dependence of T$_c$ (solid line for YBCO, Bi2212, dashed line for
LSCO) and superconducting fluctuations in Bi2212 
(circles) [\protect\onlinecite{Cor}] and (diamonds) [\protect\onlinecite{TIYM}],
LSCO (triangles) [\protect\onlinecite{Ong}], and YBCO (closed squares) 
[\protect\onlinecite{Bern}] and (open squares) [\protect\onlinecite{Mein}].}
\label{fig:38a}
\end{figure}

Above, we have presented evidence that the optimum $T_c$ occurs at a different
doping in LSCO than for the other cuprates, while the stripes terminate at
approximately the same doping for all cuprates.  Further evidence for this
can be found by comparing the superconducting 
fluctuations\cite{Cor,Ong,Bern,Mein,TIYM}, 
Fig.~\ref{fig:38a}.  In all three materials, LSCO, YBCO, and Bi2212, some 
superconducting fluctuations appear to persist in the low doping regime, to
temperatures in excess of optimal $T_c$ (although considerably less than the
pseudogap temperature).  The fluctuating superconductivity in LSCO is very
similar to that found in the other cuprates.  On the other hand, if one assumed
a universal curve of superconductivity, with $T_c/T_{c,max}$ vs $x$ the same
for all cuprates, then the scaled superconducting fluctuations in LSCO would
be at anomalously high temperatures.  It seems simpler to assume that in LSCO
long range superconducting order is anomalously suppressed (as also suggested by
Baskaran\cite{Bas2}), in which case
there is no reason for optimal $T_c$ to fall at the same doping.

In Bi2212 and YBCO, the onset of superconducting fluctuations falls close to the
`strong pseudogap' regime\cite{EmBatl,StoP}, and hence at temperatures 
substantially below the weak pseudogap temperature\cite{pseud,Ido}, making it 
unlikely that the weak pseudogap is associated with preformed pairs.  In LSCO, 
there is some confusion, since the pseudogap temperatures reported by Batlogg, 
et al.\cite{pseud} are considerably higher than those found by Ido, et 
al.\cite{Ido}.  The Ido $T^*$'s fall close to the fluctuation onset temperatures
in Fig.~\ref{fig:38a}.

The fluctuations observed in Fig.~\ref{fig:38a} are very suggestive of the
theoretical results of Fig.~\ref{fig:20c}: thus, quantum size effects enhance
the average gap on a narrow stripe, which will lead to an enhanced mean field
transition temperature in the underdoped regime.  However, the coupling between
stripes also weakens, leading to a loss of interstripe coherence, so the pairing
is only evident as enhanced fluctuations.\cite{EmK}

\section{Conclusions}

Recent experiments have provided considerable evidence for the presence of 
stripes and EPS in the cuprates, but there remain many questions of how 
universal these are, how they arise and vary with doping, and how they interact 
with superconductivity.  We have here elaborated our earlier\cite{OSP} model of
stripes driven by frustrated phase separation, in particular adducing evidence
that the doping on the charged stripes is close to $x=x_0=0.25$, and that when 
the average doping approaches this value EPS terminates.  
Moreover, near $x_{cr}=x_0/2$ there is a crossover in stripe properties: for 
$x<x_{cr}$ the charged stripes are quantum confined, for $x>x_{cr}$ the 
AFM stripes are so confined.  This model can explain the 1/8 anomaly 
($x=x_{cr}$), the anomalous Hall effect ($R_H\rightarrow 0$)\cite{Uch} for 
$x<1/8$ (charged stripes confined, hence one-dimensional), and the growing spin 
gap in YBCO for $x>1/8$\cite{OSP}.

On the important issue of the {\it structure} of a charged stripe, we have
explored a number of possibilities without coming to any final conclusions.  
While there is evidence that superconductivity lives on the charged stripes, 
there also appears to be a second instability on these stripes, which
stabilizes the stripe phase while competing with superconductivity.  We
have shown that a semiquantitative understanding can be achieved by looking at
the properties of a single doped ladder, and we have discussed how a number of
instabilities (both CDW and cuperconducting) vary with ladder width.  We showed
that strong correlation effects could lead to charged stripes with a residual 
magnetic order, introduced a simple model for White-Scalapino-like stripes, and 
found novel superconducting and CDW
instabilities associated with such stripes.  We illustrated how stripe order 
would affect ARPES spectra, both dispersions and Fermi surface maps.  Future
studies will apply the model to describing other properties of the cuprates.

Certain anomalous features of strong-correlation calculations may find an
explanation in underlying phase separation.  Thus, the vanishing of the
renormalized hopping parameter $t$ near half filling in slave boson calculations
may reflect the vanishing of the charged stripes at half filling\cite{OSP},
while the frequently-observed pinning of the VHS near the Fermi
level\cite{Surv,Vpin} is consistent with VHS-stabilized charged stripes.

Finally, the idea of a commensurability effect near 1/8 doping, leading to a
coexistence of {\it domains} for $x>1/8$, provides a simple explanation for a
large variety of experimental findings, including the saturation of the Yamada 
plot, the direct observation of domains in STM studies, and a variety of
microwave anomalies.  This may also lead to a resolution of the combined puzzle 
of magnetic neutron scattering incommensurability and the neutron resonance
peak.  A stripe model provides a natural explanation of the incommensurability
for $x\le 1/8$, including a stripe reorientation transition at the 
metal-superconductor transition near $x\sim 0.053$.  However, a band 
picture\cite{nest2} (with $E_F$ close to a VHS) provides a superior model for 
the combined, frequency-dependent incommensurability {\it cum} resonance peak 
found near optimal doping in YBCO and Bi2212.  An EPS crossover from stripes to 
domains near $x\sim 1/8$ would provide a natural explanation of these phenomena.
Interestingly, the only calculation\cite{nest1} to approximately describe the 
doping dependence is based on a slave boson model, which as noted above tends 
to mimic EPS behavior as $x\rightarrow 0$.

{\bf Acknowledgment:}  These computations were carried out using the facilities
of the Advanced Scientific Computation Center at Northesatern University 
(NU-ASCC).  Their support is gratefully acknowledged.  We had stimulating
conversations with S. Pan and J.C. Davis.  CK's research was supported in part 
by NSF Grant NSF-9711910.  Publication 776 of the Barnett Institute.

\end{document}